\documentclass{aastex61}

\usepackage{natbib}
\usepackage{graphicx,float,amsmath,bm,epsfig,epstopdf}
\shortauthors{Pal et al.}

\begin{document}

\title{Dependence of Coronal Mass Ejection Properties on their Solar Source Active Region Characteristics and Associated Flare Reconnection Flux}


\author{Sanchita Pal}
\affiliation{Center of Excellence in Space Sciences India, Indian Institute of Science Education and Research Kolkata, Mohanpur 741246, West Bengal, India}

\author{Dibyendu Nandy}
\correspondingauthor{Dibyendu Nandy}
\email{dnandi@iiserkol.ac.in}
\affiliation{Center of Excellence in Space Sciences India, Indian Institute of Science Education and Research Kolkata, Mohanpur 741246, West Bengal, India}
\affiliation{Department of Physical Sciences, Indian Institute of Science Education and Research Kolkata, Mohanpur 741246, West Bengal, India}

\author{Nandita Srivastava}
\affiliation{Udaipur Solar Observatory, Physical Research Laboratory, Udaipur 313001, Rajasthan, India}
\affiliation{Center of Excellence in Space Sciences India, Indian Institute of Science Education and Research Kolkata, Mohanpur 741246, West Bengal, India}

\author{Nat Gopalswamy}
\affiliation{NASA Goddard Space Flight Center, Greenbelt, MD 20771, USA }

\author{Suman Panda}
\affiliation{Department of Physical Sciences, Indian Institute of Science Education and Research Kolkata, Mohanpur 741246, West Bengal, India}
\affiliation{Center of Excellence in Space Sciences India, Indian Institute of Science Education and Research Kolkata, Mohanpur 741246, West Bengal, India}

\begin{abstract}
The near-Sun kinematics of coronal mass ejections (CMEs) determine the severity and arrival time of associated geomagnetic storms. We investigate the relationship between the deprojected speed and kinetic energy of CMEs and magnetic measures of their solar sources, reconnection flux of associated eruptive events and intrinsic flux rope characteristics. Our data covers the period 2010-2014 in solar cycle 24. Using vector magnetograms of source active regions we estimate the size and nonpotentiality. We compute the total magnetic reconnection flux at the source regions of CMEs using the post-eruption arcade method. By forward modeling the CMEs we find their deprojected geometric parameters and constrain their kinematics and magnetic properties. Based on an analysis of this database we report that the correlation between CME speed and their source active region size and global nonpotentiality is weak, but not negligible. We find the near-Sun velocity and kinetic energy of CMEs to be well correlated with the associated magnetic reconnection flux. We establish a statistically significant empirical relationship between the CME speed and reconnection flux that may be utilized for prediction purposes. Furthermore, we find CME kinematics to be related with the axial magnetic field intensity and relative magnetic helicity of their intrinsic flux ropes. The amount of coronal magnetic helicity shed by CMEs is found to be well correlated with their near-Sun speeds. The kinetic energy of CMEs is well correlated with their intrinsic magnetic energy density. Our results constrain processes related to the origin and propagation of CMEs and may lead to better empirical forecasting of their arrival and geoeffectiveness.

\end{abstract}


\section{Introduction} \label{sec:intro}

A Coronal mass ejection (CME) represents one of the most energetic phenomenon on the Sun, ejecting a massive amount of solar magnetized plasma (order of $10^{12}$ kg) carrying significant energy ($10^{31}-10^{33}$ erg) \citep[e.g.][]{1974JGR....79.4581G,1997cwh..conf..259H,2016GSL.....3....8G},\citep{2017SSRv..212.1159M,Green2018} in to interplanetary space. The origin of CMEs is related to the magnetic field dynamics on the solar photosphere \citep[e.g.][]{2007AAS...210.2402N}. If a CME is directed towards Earth, it may cause major geomagnetic storms depending upon its kinematics, magnetic structure and magnetic field strength at 1 AU \citep[e.g.][]{2009cwse.conf...77G},\citep{Kilpua2017}.  When a high-speed interplanetary CME (ICME) with an enhanced southward magnetic field component hits the Earth, it reconnects with the Earth's magnetosphere, enhances the ring current \citep{1998JGR...10317705K} and temporarily decreases the strength of Earth's horizontal magnetic field component. Such solar-induced magnetic storms can result in serious disruptions to satellite operations, electric power grids and communication systems. Understanding the origin of CMEs, their subsequent dynamics and developing forecasting capabilities for their arrival time and severity are therefore important challenges in the domain of solar-terrestrial physics.\par

Near-Sun kinematic properties is one of the features of CMEs that can be used to predict the intensity and onset of associated geomagnetic storms \citep{2002GeoRL..29.1287S}. In order to predict the CME arrival time at 1 AU, several empirical and physics based models constrain CME propagation through interplanetary space \citep{2001JGR...10629207G,2013SpWea..11..661G,2003JGRA..108.1445C,2003JGRA..108.1070F},\citep{2013SpWea..11..661G,2013SoPh..285..295V,2015SoPh..290.1775M,2017ApJ...837L..17T,2018ApJ...854..180D}. The models are usually based on the initial speed of CMEs. CMEs originate in closed magnetic field regions on the Sun such as active regions (ARs) \citep{2001ApJ...561..372S} and filament regions \citep{2015ApJ...806....8G}. Several studies have attempted to connect the near-Sun CME speeds and magnetic measures of their source regions \citep{2017SoPh..292...66K,2015GeoRL..42.5702T,2008ApJ...680.1516W,2002ApJ...581..694M}. \citet{2012Ge&Ae..52.1075F} studied the projected speed of 46 halo CMEs and found that the CME speed is well correlated with the average intensity of line-of-sight magnetic fields at CME associated flare onset. A recent study by \citet{GOPALSWAMY2017b} and \citet{2007ApJ...659..758Q} showed that the poloidal magnetic flux of flux rope ICMEs at 1 AU depends on the photospheric magnetic flux underlying the area swept by the flare ribbons or the post eruption arcades on one side of the polarity inversion line (defined as flare reconnection flux). Extension of these studies offer great potential for better constraining the origin and dynamics of CME flux ropes. \par
Magnetic reconnection plays an essential role at the early stage of CME dynamics. Both theoretical calculations and numerical simulations show that enhancement of CME mass acceleration is accompanied by an enhancement in the rate of magnetic reconnection at its solar source \citep{2000JGR...105.2375L, 2003ApJ...596.1341C}. Also, an observation by \citet{2004ApJ...604..900Q} revealed a temporal correlation between the reconnection rate inferred from two-ribbon flare observations and associated CME acceleration. Several previous studies attempted to compare the total flux reconnected in the CME associated flares and CME velocity and observed a strong correlation between these parameters \citep{2005ApJ...634L.121Q, 2009A&A...499..893M, GOPALSWAMY2017b}. It is well established that the acceleration phase of CMEs is synchronized with the impulsive phase of associated flares \citep{2001ApJ...559..452Z, 2003ApJ...588L..53G}. \citet{2008ApJ...673L..95T} observed a close relationship between CME acceleration and flare energy release during its impulsive phase. There exists a feedback relationship between flares and associated CMEs through magnetic reconnection that occurs in the current sheet formed below the erupting CME flux rope \citep{2010ApJ...712.1410T, 2008AnGeo..26.3089V, 2016AN....337.1002V}. This reconnection process significantly enhances the mass acceleration of the ejections as well as release energy through the accompanied two-ribbon flares \citep{2000JGR...10523153F,2000JGR...105.2375L}. These studies motivate us to explore the relationship betwen CME kinematics and the magnetic reconnection which causes the CME flux rope eruption.

CMEs are typically observed by coronagraphs which occult the photosphere of the Sun and expose the surrounding faint corona. Basic observational properties of CMEs such as their structure, propagation direction, and derived quantities such as velocity, accelerations, and mass are subject to projection effects depending on the location of the CME source region on the solar surface \citep{JGRA:JGRA17179,angeo-23-1033-2005},\citep{2007A&A...469..339V},\citep{2008JGRA..113.1104H}. The coronagraphs of the Sun-Earth Connection Coronal and Heliospheric Investigation \citep[SECCHI,][]{2008SSRv..136...67H} aboard the Solar TErrestrial RElations Observatory (STEREO) spacecrafts A \& B provide simultaneous observations of CMEs from two different viewpoints in space. Applying the forward modeling technique \citep{2006ApJ...652..763T,2009SoPh..256..111T,2011ApJS..194...33T} to CME white-light images observed from different vantage points, one can better reproduce CME morphology and dynamics. Thus deprojected CME parameters can be estimated \citep{2012SoPh..281..167B,2013JGRA..118.6858S,2013SoPh..284...47X}.\par

In this paper, we examine the size, nonpotentiality and the flare reconnection flux of CME associated flaring active regions using observations from different instruments on the Solar Dynamic Observatory \citep[SDO,][]{2012SoPh..275....3P} and connect them with CME knematics and flux properties. \citet{GOPALSWAMY2017b} studied about 50 CMEs from solar cycle 23 and their flux rope properties. Here we consider a number of CMEs from cycle 24 using a different flux rope fitting method for multi-view observations and confirm, extend and set better constraints on the relationship between CME properties and its source regions. \par

We organize this paper as follows. In Section~2 we describe the procedure of selecting CMEs and their associated solar sources and summarize the method of measuring the deprojected geometric properties of CMEs and the magnetic properties of their solar sources. In Section~3 we examine the relationship between CME kinematics with magnetic measures of their source regions as well as their intrinsic, near-Sun flux rope magnetic properties. We discuss our results in Section~4 and conclude in Section~5

\section{ Method of event selection and data}

We construct a list of 438 CMEs which have clear flux-rope morphology (determined manually) characterized by a bright front encompassing a dark cavity that surrounds a bright core and appear as a single event in each data frame of white-light synoptic movies provided by SECCHI/COR2 A \& B during solar cycle 24 (between the start of SDO mission in May 2010 and until data from both STEREO spacecrafts are available). We also identify the observed CMEs in the images obtained by the Large Angle and Spectrometric Coronagraph (LASCO) \citep{1995SoPh..162..357B} telescope's C2 and C3 on board Solar and Heliospheric Observatory \citep[SOHO,][]{ 1995SoPh..162....1D}. The corresponding solar source location of the CMEs were determined using SDO's Atmospheric Imaging Assembly (AIA) \citep{2012SoPh..275...17L} images at 193 \AA\ and SECCHI's Extreme Ultraviolet Imager (EUVI) data at 195 \AA. From the list of selected events we isolate those which originated on the Earth facing side of the Sun. In our study, we consider the source ARs within $\pm 45 ^{\circ}$ longitude from the disk center to avoid projection effects in magnetogram observations of ARs. We further short list the events by the requirement that their source regions have been identified by NOAA and that their vector magnetograms exist from Helioseismic Magnetic Imager observations \citep [HMI,][]{2012SoPh..275..207S} on board SDO. This careful manual selection method leaves only 36 CMEs for our study.  \par

The flux rope structure of the identified CMEs allows us to apply the Graduated Cylindrical Shell (GCS) forward modeling technique developed by \citet{2006ApJ...652..763T}. The GCS model helps derive the deprojected parameters of CMEs from projected white-light images \citep[e.g.][]{2010ApJ...722.1762L,2010ApJ...717L.159P,2011ApJ...733L..23V}. The six geometric parameters, which model the flux rope CMEs are the propagation longitude ($\phi$), latitude ($\theta$), aspect ratio ($\kappa$), tilt angle ($\gamma$) between the source region neutral line and the equator, the half angular width between the legs ($\alpha$) and the height ($h$) of the CME leading edge (see Figure 1 of \citet{2006ApJ...652..763T}). By adjusting these six parameters manually, we try to achieve the best match between the model CMEs and the observed CMEs in LASCO and STEREO coronagraphs. In Figure 1, we show an example of GCS model fitting result. The model is applied to COR2 A \& B and calibrated (Level 1) LASCO C3 base difference white-light CME images. The COR2 images are used after being processed by the standard routines (secchi\_prep) available in SolarSoft. For a well fitted CME, we obtain the CME speed by tracking its leading edge until it reaches the edge of the field of views (FOVs) of the coronagraphs. Some of the observed CMEs become faint before reaching the edges of the FOVs of the coronagraphs. The deprojected propagation speed of CME ($V_{gcs}$) we quote here is obtained by linear fitting of the height-time measurement of CME leading edges propagating within the FOVs of the coronagraphs. \par

To obtain the magnetic properties of source ARs, we use Space-Weather HMI AR Patch (SHARP) data series (hmi.Sharp\textunderscore cea\textunderscore 720s) and full disk HMI vector magnetogram data series (hmi.B\textunderscore720s) along with the AIA 193 \AA\ data. The hmi.B\textunderscore720s data series provides vector field information in the form of field strength, inclination and azimuth in plane-of-sky co-ordinates \citep{2014SoPh..289.3483H}. We perform a co-ordinate transformation and decompose the magnetic field vectors into r (radial distance), $\theta$ (polar angle), and $\phi$ (azimuthal angle) components in spherical co-ordinates \citep{2013arXiv1309.2392S}. To derive the vector magnetic field components we use HMI pipeline codes publicly available in the SDO webpage. In our data set, we find many ARs identified with different NOAA numbers although they are magnetically connected. Therefore, we use SHARP vector magnetograms (as each AR patch includes single or multiple connected ARs) to measure the global magnetic parameters of source ARs.

\begin{figure}[!tbp]
  \centering
   \includegraphics[width=0.7\textwidth]{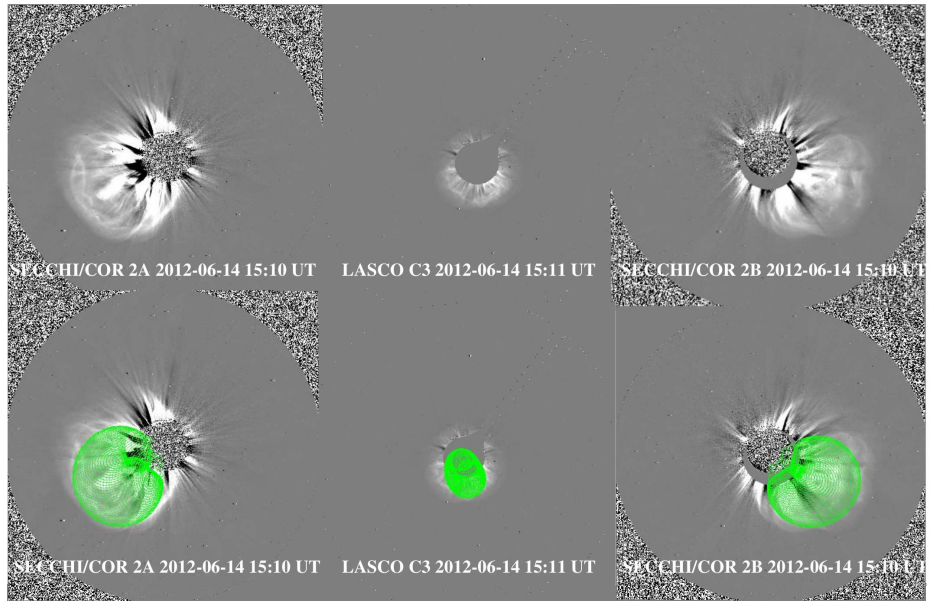}
  \caption{Forward modeling of white-light images of CME (observed on 14 June, 2012) with the GCS model. Top three panels (left-to-right) represent CME white-light images observed by STEREO B, LASCO C3, and STEREO A, respectively. Bottom three panels show CME with GCS wire frame (in green symbol) overlaid on top. The fitting results the deprojected geometric parameters of the CME as, $\phi= 89.07^{\circ}$, and $\theta= -32^{\circ}$ (in Carrington co-ordinate), $\gamma= -67^{\circ}$, $\kappa =0.58$, $\alpha= 23^{\circ}$, and $h= 10.5 R_{s}$ }

\end{figure}

\subsection{Magnetic properties of ARs and CMEs}

In this section, we discuss the magnetic properties of ARs and describe the methods used to measure their properties. Guided by widely utilized AR characteristics in the community in this context, we consider a few relevant AR parameters for our study. We determine the total unsigned magnetic flux as a proxy of AR size. We also determine the AR nonpotentialty through estimates of three different proxies -- total unsigned vertical current, total photospheric magnetic free energy density, and length of the strong field neutral line. We further compute the magnetic reconnection flux in the low corona associated with each event by utilizing the fact that post eruption arcades (PEAs) map out the reconnection region leading to formation of flux ropes during solar eruptive events. \citep{2007ApJ...659..758Q,2007ApJ...669..621L,2014ApJ...793...53H,2017aSoPh..292...65G}. We obtain the magnetic properties of CME flux rope following the Flux Rope from Eruption Data (FRED) technique that combines the reconnection flux with geometrical flux rope properties \citep{GOPALSWAMY2017b,2017carXiv170903160G,0004-637X-851-2-123}.
.
\begin{figure}
\centering
  \includegraphics[width=0.7\textwidth]{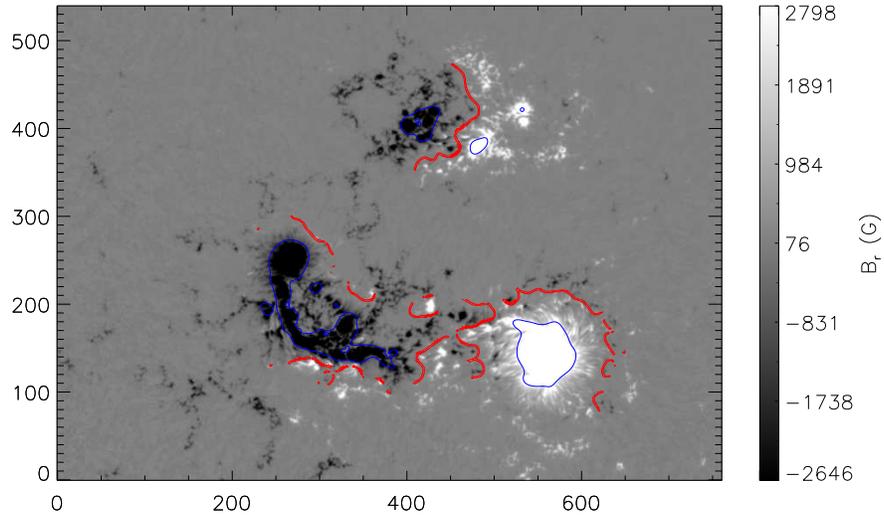}
   \caption{SHARP vector magnetogram of AR 11504 on 14 June 2012 from which a CME erupted at 12:48 UT. Blue contours define the region above the disambiguation noise threshold ($B_r\approx 150$ G, CONF\textunderscore DISAMBIG= 90). The red lines denote the strong field neutral lines associated  with the magnetic field distribution. The vertical, gray color bar shown on the right depicts the values of $B_r$. The maximum and minimum $B_r$ of the AR are respectively, 2798 G and -2645 G.}
\end{figure}

\subsection{Total unsigned magnetic flux}

The total unsigned magnetic flux ($\phi_{AR}$) of an AR is calculated by integrating the radial magnetic field component ($B_r$) over the high-confidence region within the HARP. Here the high-confidence region is defined by cluster of pixels above the disambiguation noise threshold (where the confidence in disambiguation, CONF\textunderscore DISAMBIG is equal to 90; see Table A.5 of \citet{2014SoPh..289.3549B}). Thus, $\phi_{AR}$ is defined by,
\begin{equation}
\phi_{AR}=\int \mid B_{r}\mid dA
\end{equation}

Here each pixel area is defined by A (= $0.5"\times 0.5"$).
In figure 2, we display an example of a SHARP vector magnetogram of AR NOAA 11504 located at S17E06, where, the blue contours enclose regions with $B_r$ values greater than the noise threshold.

\subsection{Total vertical current}
The vertical current density ($J_z$) is measured using Ampere's current law which gives,
\begin{equation}
J_z=\frac{1}{\mu}(\frac{\partial B_y}{\partial x}-\frac{\partial B_x}{\partial y}),
\end{equation}
Where $B_x$ and $B_y$ are the observed horizontal components of AR magnetic field and $\mu$ is the magnetic permiability.
The total unsigned vertical current ($I_{tot}$) is computed by integrating $J_z$ over all pixels above the noise threshold (CONF\textunderscore DISAMBIG= 90).

\subsection{Total photospheric free magnetic energy density ($\rho_{tot}$)}
 \citet{1996ApJ...456..861W} define the density of the free magnetic energy ($\rho_e$) in terms of observed magnetic field ($B_{obs}$) and potential magnetic field ($B_p$) components obtained from vector magnetogram. The formula that is used to calculate this measure is,
\begin{equation}
\rho_e=\frac{(B_{obs}-B_p)^2}{8\pi}
\end{equation}
Now $\rho_{tot}$ is measured by integrating $\rho_e$ over all the pixels above the noise threshold.

\subsection{Length of strong field neutral line }

 The length of the strong field neutral line, $L_{nl}$ is formulated as,
 \begin{equation}
L_{nl}=\int_{B_{pt} > 150 G} dl
\end{equation}
 Here the integration involves all neutral line increments $dl$ on which the transverse potential magnetic field component ($B_{pt}$) of the vector magnetogram is greater than 150 G \citep{2008ApJ...689.1433F,2011SpWea...9.4003F}. Also $dl$ separates opposite polarities of $B_r$ of at least 20 G \citep{2008ApJ...689.1433F}. We calculate $B_{pt}$ from $B_r$, where $B_r$ is greater than the noise threshold. In Figure 2 we indicate the locations of neutral lines (in red) on which the transverse potential field is greater than 150 G.

\subsection{Magnetic reconnection flux}
To measure the magnetic reconnection flux ($\phi_{RC}$), we use the PEA technique proposed by \citet{2017aSoPh..292...65G}. In our study, we identify 33 out of 36 CMEs for which post-eruption loops are clearly observed in AIA 193 \AA\ images. We mark the foot prints of PEAs on AIA 193 \AA\ images and define the area under the PEAs by creating a polygon connecting the marked foot prints. We then overlay the polygon on the differential-rotation corrected full disk HMI vector magnetogram obtained $\approx 30$ minutes before the onset of the eruption and integrate the absolute value of $B_r$ in all the pixels within the polygon. The resulting $\phi_{RC}$ is half of the total flux through the polygon.
Therefore, $\phi_{RC}$ is defined by,
\begin{equation}
\phi_{RC}=\frac{1}{2}\int_{PEA} \mid B_{r}\mid dA
\end{equation}
In figure 3 (a) and (b) we show NOAA AR11504 in 193 \AA\ (from the AIA instrument) and its vector magnetogram, respectively. The red dashed lines on both the images define the PEA foot prints.

\begin{figure}[!tbp]
  \centering
  \includegraphics[width=0.7\textwidth]{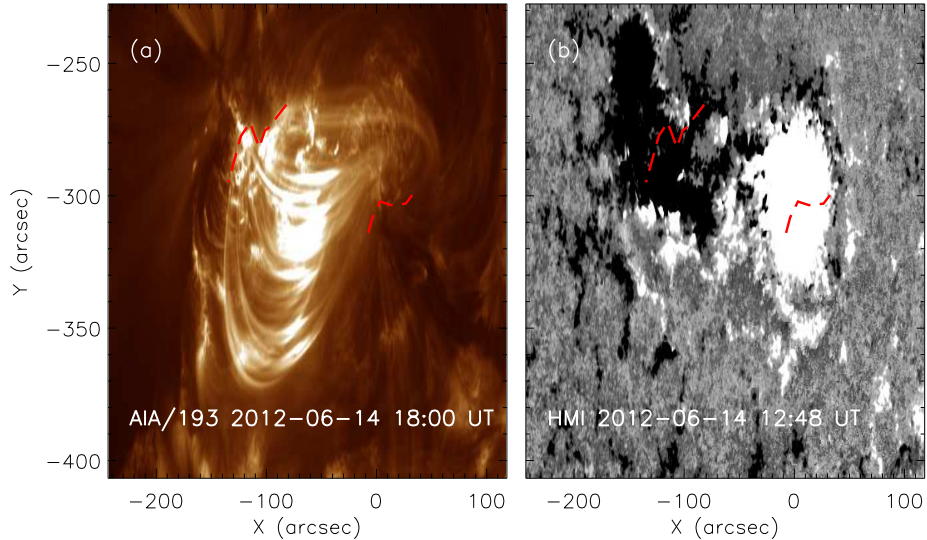}
  \caption{Post eruption arcade (PEA) and corresponding vector magnetogram associated with the 14 June, 2012 CME. (a) SDO/AIA/193 \AA\ image of PEA observed in low corona at 18:00 UT. (b) HMI vector magnetogram of AR 11504 (solar source of the observed CME) at 12:48 UT. The red dashed lines in (a) and (b) represent the PEA foot prints. The $\phi_{RC}$ associated with the arcade is 7.45$\times 10^{21}$ Mx.}
\end{figure}

\subsection{Relative magnetic helicity }

The relative magnetic helicity, $H_m$, is derived by subtracting the reference magnetic field ($B_{ref}$) helicity from the magnetic helicity ($H$) of a field $B$ within a volume V \citep{berger_field_1984} and is given by
\begin{equation}
H_m= \int_{V} \bm{A\cdot B}\ dV -\int_{V} \bm{A}_{ref}\cdot \bm{B}_{ref}\ dV
\end{equation}
Here $A$ is the vector potential. For a CME with flux rope structure, $\bm{B_{ref}}=B_z \hat{z}$ and $\bm{B}= B_\phi \hat{\phi}+B_z\hat{z}$, where Bz is the axial magnetic field component and $B_{\phi}$ is the azimuthal magnetic field component of a cylindrical flux rope. The magnetic field components are derived using Lundquist's constant-$\alpha$ force-free field solution in cylindrical coordinates \citep{1990JGR....9511957L}. Using $\bm{A}= \bm{B}/\alpha$ we calculate the magnetic helicity of a CME flux rope as \citep{2003JGRA..108.1362D,2002A&A...382..650D,2000ApJ...539..944D},
\begin{equation}
H_m= 4\pi L \int_0^{R_{0}} A_{\phi}B_{\phi}\ rdr\approx 0.7\ B_{0}^{2}R_{0}^{3}L
\end{equation}
Here $R_0$ is the radius of the circular annulus of CME at it's leading edge point. It is defined by $R_{0}=h/(1+(1/\kappa)$ estimated using equation (1) of \citet{2006ApJ...652..763T}. $L$ is the length of CME flux rope approximated as $L= 2h_{leg}+(\pi/2+\alpha)(h-h_{leg}/cos\alpha)-2R_{\odot}$ \citep{0004-637X-851-2-123}, where $h_{leg}$ is the height of the CME flux rope legs (see equation (3) of \citet{2006ApJ...652..763T}) and $(\pi/2+\alpha)$ is in radian. $B_0$ is the axial magnetic field strength of the CME defined by $B_{cme}= \phi_{p}x_{01}/LR_{0}$ (assuming a force-free CME flux rope). Here $\phi_p$ is the azimuthal flux of CME which is approximately equal to $\phi_{RC}$ and $x_{01}= 2.4048$ is the zeroth order Bessel function.

\section{Analysis and Results}

In this section, we analyze the relationship between CME kinematic properties, magnetic properties of their solar source regions, reconnection flux and associated flux rope characteristics. The inferred parameters are summarized in Table 1 which lists 36 CMEs and their properties along with the associated solar source information. The event numbers are shown in column 1. In column 2, we mention the dates and times when the CMEs first appear in the LASCO coronagraphs (CDAW LASCO CME catalog \citep[\url{http://cdaw.gsfc.nasa.gov/CME_list/},][]{2004JGRA..109.7105Y,2009EM&P..104..295G} ). Column 3 shows the NOAA numbers of the CME associated source ARs. Column 4-9 represent the magnetic information ($\phi_{AR}$, $I_{tot}$, $\rho_{tot}$, $L_{nl}$, $\phi_{RC}^p$, $\phi_{RC}^r$) of the identified ARs. Column 10 lists the mass of corresponding CMEs ($M_{cme}$) collected from LASCO CME catalog. Column 11 and 12 list $\alpha$ and $V_{gcs}$ of CMEs. Column 13 and 14 represent the magnetic properties of CMEs -- $B_{cme}$, and $H_m$.\par

\begin{deluxetable}{cccccccccccccc}

\rotate



\tabletypesize{\scriptsize}






\tablecaption{Properties of selected CMEs and associated source region information}

\tablenum{1}

\tablehead{\colhead{Event} & \colhead{Date \& time} & \colhead{NOAA} & \colhead{$\phi_{AR}$} & \colhead{$I_{tot}$} & \colhead{$\rho_{tot}$} & \colhead{$L_{nl}$} & \colhead{$\phi_{RC}^p$} & \colhead{$\phi_{RC}^{r(1)}$} & \colhead{$M_{cme}$} & \colhead{$\alpha$} & \colhead{$V_{gcs}$} & \colhead{$B_{cme}$} & \colhead{$H_m$} \\
\colhead{number} & \colhead{(DD-MM-YYYY hh:mm UT)} & \colhead{number} & \colhead{($10^{22}$ Mx)} & \colhead{($10^{14}$ A)} & \colhead{($10^{24}$ erg cm$^{-1}$)} & \colhead{($10^{5}$ km)} & \colhead{($10^{21}$ Mx)} & \colhead{($10^{21}$ Mx)} & \colhead{($10^{15}$ gm)} & \colhead{($^{\circ}$)} & \colhead{(km s$^{-1}$)} & \colhead{(mG)} & \colhead{($10^{42}$ Mx$^2$)} }

\startdata
1$^b$ & 01-08-2010 13:42 & 11092 & 1.28 & 0.533 & 0.4 & 0.306 & 9.36 & 2.96 & - & 23.20 & 1260 & 62.59 & 86.30 \\
2$^b$ & 07-08-2010 18:36 & 11093 & 0.89 & 0.543 & 0.21 & 0.0288 & 1.58 & 4.75 & - & 14.81 & 779 & 14.92 & 1.87 \\
3 & 14-02-2011 18:24 & 11158 & 2.50 & 1.39 & 0.83 & 5.63 & 4.54 & - & 0.86 & 22.36 & 359 & 51.44 & 12.40 \\
4 & 15-02-2011 02:24 & 11158 & 2.69 & 1.55 & 0.85 & 5.15 & 10.4 & 11.6 & 4.3 & 28.51 & 868 & 119.13 & 62.90 \\
5 & 01-06-2011 18:36 & 11226 & 2.81 & 1.73 & 0.36 & 3.28 & 1.49 & 2.2 & 1.8 & 22.64 & 527 & 20.29 & 1.09 \\
6 & 02-06-2011 08:12 & 11227 & 2.39 & 1.67 & 0.34 & 3.1 & 1.81 & 1.7 & 1.4 & 17.33 & 1176.4 & 42.63 & 0.96 \\
7 & 21-06-2011 03:16 & 11236 & 1.98 & 1.46 & 0.41 & 1.82 & 6.1 & 1.13 & 6.2 & 26.55 & 970 & 72.40 & 21.10 \\
8 & 09-07-2011 00:48 & 11247 & 0.16 & 0.14 & 0.01 & 0.356 & 3.54 & - & 1.8 & 23.20 & 861 & 33.56 & 8.89 \\
9 & 03-08-2011 14:00 & 11261 & 2.42 & 1.69 & 0.49 & 3.63 & 4.4 & 7.61 & 8.7 & 17.90 & 1228 & 55.26 & 10.70 \\
10 & 04-08-2011 04:12 & 11261 & 2.56 & 1.81 & 0.44 & 3.74 & 5.58 & 8.26 & 11 & 24.87 & 1737 & 69.37 & 17.00 \\
11 & 06-09-2011 23:05 & 11283 & 1.73 & 1.24 & 0.33 & 2.6 & 5.59 & 5.92 & 15 & 35.50 & 900 & 52.40 & 20.90 \\
12 & 07-09-2011 23:05 & 11283 & 1.89 & 1.31 & 0.31 & 2.5 & 8.44 & 7.98 & 1.1 & 15.93 & 914 & 79.43 & 53.30 \\
13$^a$ & 24-09-2011 19:36 & 11302 & 5.73 & 2.35 & 1.82 & 8.47 & - & - & 3.1 & 21.24 & 944.46 & - & - \\
14 & 09-11-2011 13:36 & 11343 & 1.06 & 0.475 & 0.19 & 0.418 & 5.4 & 6.36 & 14 & 35.78 & 1285 & 29.02 & 28.30 \\
15 & 26-12-2011 11:48 & 11384 & 2.08 & 1.27 & 0.6 & 2.06 & 1.95 & 1.09 & 4.3 & 6.98 & 777 & 21.79 & 2.46 \\
16 & 19-01-2012 14:36 & 11402 & 7.01 & 4.02 & 1.32 & 7.07 & 10.4 & 4.56 & 19 & 25.50 & 1069 & 119.83 & 63.30 \\
17 & 23-01-2012 03:12 & 11402 & 7.10 & 4.39 & 0.89 & 7.26 & 14.3 & 17.2 & 5.3 & 43.60 & 1916 & 116.40 & 144.00 \\
18$^a$ & 06-06-2012 20:36 & 11494 & 1.15 & 0.74 & 0.3 & 2.58 & - & 2.05 & 2.6 & 13.13 & 569.4 & - & - \\
19 & 14-06-2012 14:12 & 11504 & 3.75 & 1.96 & 1.19 & 9.62 & 7.45 & 3.88 & 12 & 23.00 & 1146 & 56.40 & 48.00 \\
20 & 02-07-2012 20:24 & 11515 & 3.62 & 2.29 & 0.91 & 7.78 & 4.78 & 4.78 & 8.6 & 19.85 & 715 & 58.23 & 12.90 \\
21 & 03-07-2012 00:48 & 11515 & 4.45 & 4.54 & 1.01 & 0.283 & 2.44 & 3.78 & 3 & 12.90 & 409 & 36.59 & 2.76 \\
22 & 12-07-2012 16:48 & 11520 & 9.04 & 5.26 & 2.28 & 13.7 & 13.3 & 8.64 & 6.9 & 26.00 & 1700 & 129.30 & 103.00 \\
23 & 14-08-2012 01:25 & 11543 & 1.47 & 0.974 & 0.43 & 3.34 & 1.3 & 1.04 & 1.8 & 15.40 & 457 & 15.25 & 1.01 \\
24 & 28-09-2012 00:12 & 11577 & 2.41 & 1.75 & 0.24 & 2.15 & 2.81 & 2.33 & 9.2 & 30.00 & 1229.16 & 24.43 & 5.79 \\
25 & 20-11-2012 12:00 & 11616 & 1.57 & 1.25 & 0.21 & 2.01 & 3.09 & - & 8.4 & 32.70 & 719 & 49.21 & 4.08 \\
26 & 13-03-2013 00:24 & 11692 & 2.56 & 1.19 & 0.49 & 1.67 & 4.79 & 1.64 & 4.2 & 23.00 & 680.5 & 48.88 & 15.20 \\
27 & 15-03-2013 07:12 & 11692 & 1.71 & 1.11 & 0.43 & 1.74 & 4.75 & - & 13 & 25.16 & 1354.4 & 64.23 & 11.40 \\
28 & 11-04-2013 07:24 & 11719 & 1.83 & 1.55 & 0.25 & 2.45 & 5.04 & 4.5 & 22 & 36.33 & 1063 & 69.35 & 12.30 \\
29 & 07-05-2013 09:36 & 11734 & 4.54 & 2.42 & 0.78 & 4.33 & 1.3 & 1.15 & 4.3 & 12.60 & 361 & 18.54 & 0.83 \\
30 & 28-06-2013 02:00 & 11777 & 0.89 & 0.573 & 0.2 & 1.07 & 1.92 & 1.02 & 6.6 & 21.80 & 1069 & 38.29 & 1.28 \\
31 & 07-08-2013 18:24 & 11810 & 0.58 & 0.418 & 0.03 & 0.356 & 2.29 & - & 3.1 & 23.48 & 521 & 21.72 & 3.71 \\
32 & 12-08-2013 12:00 & 11817 & 1.81 & 0.799 & 0.27 & 1.94 & 2.75 & 3.46 & 3.1 & 19.30 & 395.8 & 49.87 & 2.88 \\
33 & 17-08-2013 19:12 & 11818 & 1.55 & 0.99 & 0.41 & 2.05 & 6.09 & 6.1 & 12 & 25.43 & 986 & 73.71 & 20.70 \\
34$^a$ & 26-10-2013 12:48 & 11877 & 3.33 & 2.08 & 0.76 & 4.32 & - & 0.8 & 3.3 & 20.12 & 472 & - & - \\
35 & 07-01-2014 18:24 & 11944 & 8.38 & 4.78 & 2.82 & 12.8 & 10.9 & 11.6 & 22 & 31.30 & 2187.8 & 124.16 & 68.70 \\
36 & 29-03-2014 18:12 & 12017 & 1.30 & 0.931 & 0.18 & 1.36 & 5 & 4.94 & 5 & 25.16 & 673.6 & 52.79 & 15.90 \\
\enddata


\tablecomments{$^a$ Events with undetected PEAs. \\ $^b$Events with unavailable mass information in LASCO CME catalog. \\ $^{(1)}$Data collected from RibbonDB catalog.}


\end{deluxetable}

\subsection{Magnetic properties of ARs versus associated CME speeds}

In Figure 4 we plot the deprojected speed of CMEs versus the unsigned magnetic flux and nonpotential parameters ($I_{tot}$, $\rho_{tot}$, and $L_{nl}$) of their progenitor ARs. We perform a correlation analysis and estimate the linear correlation coefficients ($r$) along with the confidence levels defined by (1-P-value). The P-value refers to the probability value of finding a result in a statistical study when the null hypothesis is true. We mention $r$ and (1-P-value) in each of the plots of Figure 4. The confidence The correlation analysis implies a weak positive correlation between CME speeds and each of the AR magnetic parameters. The similarity of the correlation coefficients imply that the analyzed AR parameters are also inter-related, plausibly, through their dependence on AR size.

Our result is in agreement with numerical simulations which suggest that an AR can produce both fast and slow CMEs but the larger and more complex (nonpotential) ones produce the fastest CMEs \citep{2007AN....328..743T}. Often, it is only a small part of a large AR that is involved in an eruption \citep{2015GeoRL..42.5702T}. Therefore, a single eruption is not enough to release the total free energy stored in ARs. Depending upon the release of free energy in each eruption, the associated CME speed may vary from slow to fast. So, complex ARs are capable of producing single or multiple eruptions and one should not necessarily expect a strong correlation between the CME speeds associated with individual events and source AR properties.

\begin{figure}
\centering
  \includegraphics[width=0.8\textwidth]{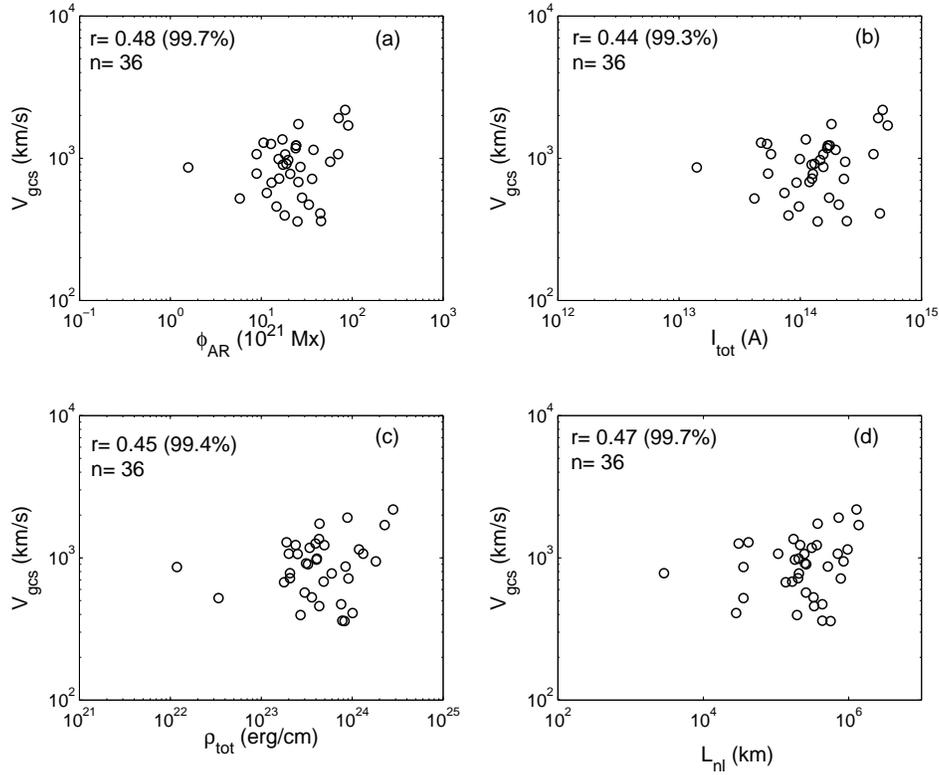}
   \caption{Scatter plots between $V_{gcs}$ and (a) $\phi_{AR}$, (b) $I_{tot}$, (c) $\rho_{tot}$, and (d) $L_{nl}$ in our dataset. The Pearson's linear correlation coefficients ($r$), confidence levels and the number of events ($n$) are mentioned in each of the plots.}
\end{figure}

\subsubsection{$\phi_{RC}$ of ARs versus properties of CMEs}

Several investigations show that reconnection of coronal field lines during eruptive events like flare results in the formation of PEAs as well as flux ropes.  In this section we identify the AR segments involved in eruptions using PEAs formed due to the flare reconnection process. We estimate the reconnection flux ($\phi_{RC}$) of these segments and analyze their influence on CME kinematics. In Figure 5, we plot $\phi_{RC}$ versus $V_{gcs}$. The data points marked by `o' (black) and `+' (red) in the plot denote $\phi_{RC}$ measured using PEAs (referred as $\phi_{RC}^p$) and ribbons (referred as $\phi_{RC}^r$), respectively. We acquire $\phi_{RC}^r$ from the RibbonDB catalog \citep[\url{http://solarmuri.ssl.berkeley.edu/~kazachenko/RibbonDB/},][]{2017ApJ...845...49K}. The catalog contains the active region and flare-ribbon properties of 3137 flares of GOES class C1.0 and larger located within 45 degrees from the central meridian and observed by SDO from April 2010 until April 2016. We find a significant positive correlation between $V_{gcs}$ and $\phi_{RC}$ for both $\phi_{RC}^p$ and $\phi_{RC}^r$ (which are similar in their strength). The correlation coefficients are respectively 0.66 and 0.68 at 99.99\% confidence level. The correlations are quite similar because the $\phi_{RC}$ for both the cases agree quite well (as was first shown by \citet{2017aSoPh..292...65G}). The correlation coefficients are lower than that reported by \citet{2005ApJ...634L.121Q} for 13 events and \citet{2009A&A...499..893M} for 21 events but similar to that of \citet{GOPALSWAMY2017b} for 48 events of solar cycle 23. The linear least-squares fits to the relationships yield the regression equations,
\begin{equation}
V_{gcs}= 327\phi_{RC}^{p\ 0.69} {km} {s^{-1}},
\end{equation}
and
\begin{equation}
 V_{gcs}=430 \phi_{RC}^{r\ 0.58} {km} {s^{-1}},
\end{equation}
respectively. Here $\phi_{RC}$ is in unit of 10$^{21}$ Mx.

\begin{figure}
\centering
  \includegraphics[width=0.5\textwidth]{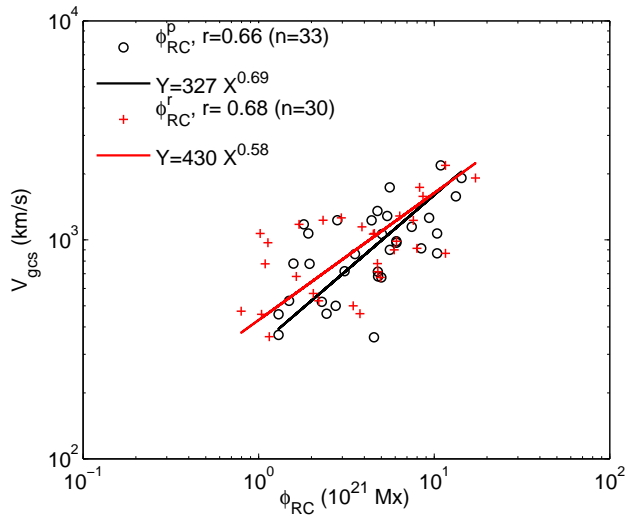}
   \caption{Scatter plot between $\phi_{RC}$ and associated $V_{gcs}$. The markers `o' and `+' denote $\phi_{RC}^p$ and $\phi_{RC}^r$, respectively. The correlation coefficients ($r$) and number of events ($n$) corresponding to both $\phi_{RC}^p$ and $\phi_{RC}^r$ are shown in the plot.  The black and red solid lines are the least-squares fits to the $\phi_{RC}^p$-$V_{gcs}$ and  $\phi_{RC}^r$-$V_{gcs}$ pairs. The regression line equation for each solid line is depicted in the figure.}
\end{figure}

We analyze the relationship between $\phi_{RC}$ and kinetic energy of the resulting CMEs. Initially, we use CME masses ($M_{cme}$) listed in CDAW LASCO CME catalog and $V_{gcs}$ to calculate the kinetic energy of CMEs ($KE_{cme}^L$). In Figure 6.(a) we show the correlation between $\phi_{RC}$ and $KE_{cme}^L$. We find a weak positive correlation with a correlation coefficient of 0.44 which is greater than the Pearson's critical correlation coefficient ($r_c$= 0.306) at 95\% confidence level. It is well known that mass of wide CMEs measured using SOHO/LASCO white-light images suffers from serious projection effects. To estimate the true masses ($M_{cme}^t$) of CMEs, we use CME angular widths ($AW$s) in the equation $log M_{cme}^t= 12.6\ log AW$ \citep{2005JGRA..11012S07G}. The positive correlation (r= 0.56 at 99 \% confidence level) between $AW$ and $\phi_{RC}$ (shown in Figure 6.(b)) statistically confirms that CME's final angular width can be estimated from the magnetic flux under the flare arcade \citep{2007ApJ...668.1221M} which is equal to $\phi_{RC}$ in our case. Since $\phi_{RC}$ is proportional to $AW$, we do expect a better correlation between $\phi_{RC}$ and $M_{cme}^t$ which further provides a good positive correlation between $\phi_{RC}$ and kinetic energy of associated CMEs ($KE_{cme}$) measured using mass, $M_{cme}^t$ and $V_{gcs}$. In Figure 6.(c), we show the correlation between $\phi_{RC}$  and $KE_{cme}$. We find $r$= 0.68 at 99.9\% confidence level and derive the regression equations of the least-squares fits (see Fig. 6). The correlation coefficient and the slope of fitted regression line are very similar to that obtained by \citet{GOPALSWAMY2017b} for cycle 23. The significant correlation between $KE_{cme}$ and $\phi_{RC}$ confirms that $\phi_{RC}$ is a good indicator of CME kinetic energy. The CME acceleration is mainly driven by the Lorentz force component representing the magnetic pressure gradient and a diamagnetic effect that comes from the induced eddy current at the solar surface \citep{Green2018, 2015SoPh..290.3457S}. The acceleration is limited by the inductive decay of the electric current that implies the decrease of Lorentz force and the free energy contained in the system \citep{2016AN....337.1002V,2010ApJ...717.1105C}. In our study, the positive correlation found between $KE_{cme}$ and $\phi_{RC}$ suggests that the reconnected field lines cause a rapid energy deposition in corresponding CME flux ropes. Here $\phi_{RC}$ serves as a proxy for reconnected magnetic field intensity.

\begin{figure}
\centering
  \includegraphics[width=1.0\textwidth]{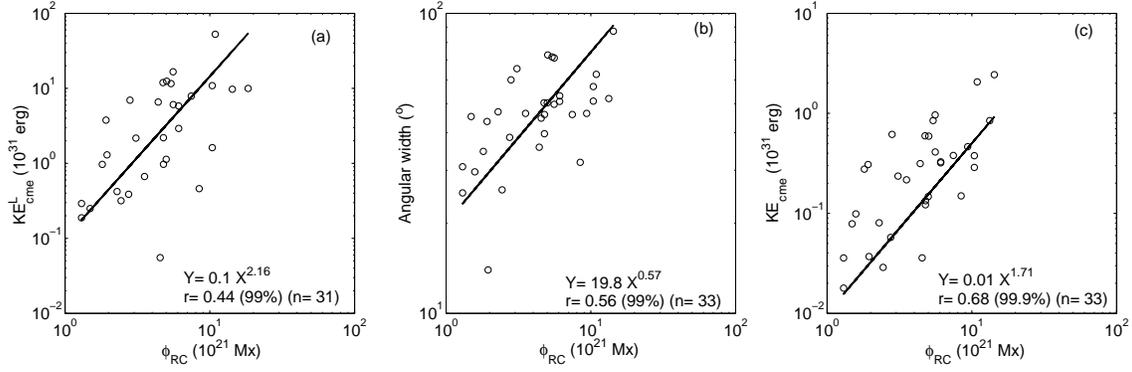}
   \caption{Scatter plots between $\phi_{RC}$ and (a) $KE_{cme}^L$, (b) angular width ($AW$), and (c) $KE_{cme}$. The correlation coefficients ($r$) mentioned in each plot suggest a significant positive correlation between each of the CME parameters and $\phi_{RC}$. The solid black lines are the least-squares fits to the plots. The regression equations are mentioned in the plots.}
\end{figure}

\begin{figure}
\centering
  \includegraphics[width=1.0\textwidth]{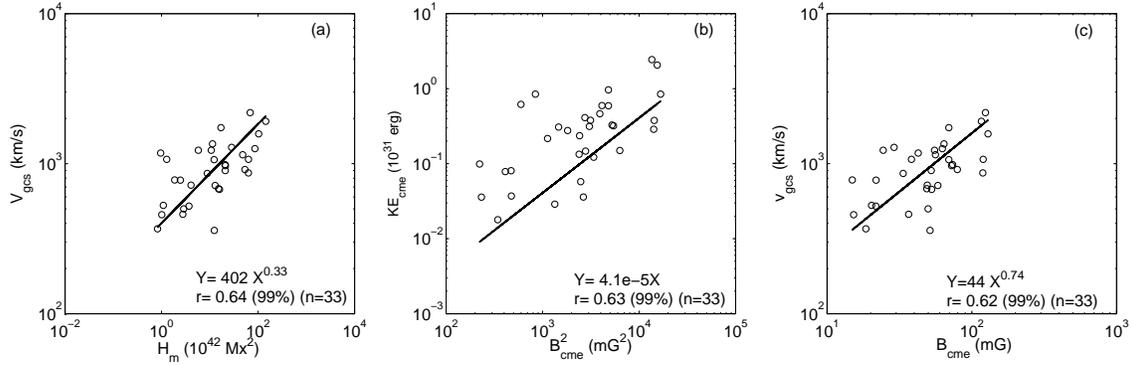}
   \caption{Scatter plots between (a) $V_{gcs}$ and $H_m$, (b) $KE_{cme}$ and $B_{cme}^2$, and (c) $V_{gcs}$ and $B_{cme}$ of CME flux ropes at 10 $R_s$. The straight line in each plot shows the linear least-squares fit to the data. The correlation coefficients ($r$) along with the equations of the regression lines are mentioned in each plot.}
\end{figure}

\subsection{Kinetic properties versus magnetic properties of CMEs}

In Figure 7.(a), (b) and (c), we study the relationship between CME kinematics (velocity and kinetic energy) and intrinsic CME magnetic properties ($B_{cme}$, magnetic pressure ($B_{cme}^2$), and $H_m$). According to the FRED technique, the axial magnetic field strength $B_{cme}$ of the flux rope depends on its geometric parameters (from the GCS model) and $\phi_{RC}$ under the assumption that the CME flux rope is force-free \citep{GOPALSWAMY2017b,2017carXiv170903160G}. We derive $B_{cme}$ as well as $H_m$ at 10 $R_s$ from $\phi_{RC}^p$ and statistically establish a positive correlation between $H_m$ and $V_{gcs}$ (shown in Figure 7.(a)), $B_{cme}^2$ and $KE_{cme}$ (shown in Figure 7.(b)) as well as $B_{cme}$ and $V_{gcs}$ (shown in Figure 7.(c)). The correlation coefficients are respectively 0.64, 0.63, and 0.62 at 99\% confidence level. The correlations suggest that CME flux ropes with higher magnetic field strength and helicities tend to have higher speeds and energies -- which is not unexpected because the CME kinematics is governed by the free magnetic energy contained in its current carrying sheared and twisted magnetic field structure \citep{2008AnGeo..26.3089V}. We find that at a radial distance ($R_{rad}$) of 10 $R_s$, the average magnetic pressure of a CME flux rope is an order of magnitude greater than the background magnetic pressure ($B_{bg}^2$) computed using $B_{bg}(R_{rad})= 0.356R_{rad}^{-1.28}$ for an adiabatic index of 5.3 \citep{2011ApJ...736L..17G}. This plausibly explains our observations that CME flux ropes with large magnetic content expands faster through the interplanetary medium \citep{2014GeoRL..41.2673G}.

\section{Discussion}

 We investigate the dependence of the initial speed of CMEs on the magnetic properties of their source ARs, reconnection flux of associated eruptive event, and the intrinsic magnetic characteristics of the CME flux rope. We measure the proxies of AR size (i.e., $\phi_{AR}$), nonpotentiality (i.e., $I_{tot}$, $\rho_{tot}$, and $L_{nl}$ ) and find a weak positive correlation ($r\approx 0.5$) between CME speed and the measured AR parameters. \citet{2017darXiv170903165G} pointed out that the magnetic reconnection flux ($\phi_{RC}$) is typically smaller than the total unsigned magnetic flux ($\phi_{AR}$) of an AR. For our events, we find the average ratio of $\phi_{RC}$ and $\phi_{AR}$ as 0.3. The value of $\phi_{RC}$/$\phi_{AR}$ suggests that only a smaller section of the active region is involved in a given eruption. This fact might be the main reason for a weak positive correlation between CME speeds and associated global, source AR parameters.

 \citet{2015GeoRL..42.5702T} studied 189 CMEs to investigate the relationship between CME speed and their sources. The study did not find any correlation between the projected CME speed and the global area and nonpotentiality of their sources. \citet{2017SoPh..292...66K} studied 22 CMEs of solar cycle 24 and examined the relationship between the CME speed, calculated from the triangulation method and the average magnetic helicity injection rate ($|\dot{H_{avg}}|$) and the total unsigned magnetic flux [$\phi(t_f)$]. They classified the selected events into two groups depending on the sign of injected helicity in the CME-productive ARs. For group A (containing 16 CMEs for which the helicity injection in the source ARs had a monotonically increasing pattern with one sign of helicity), the correlation coefficient for CME speed and $|\dot{H_{avg}}|$ was found to be 0.31, and for CME speed and $\phi(t_f)$ it was 0.17. Whereas, for group B (containing only 6 CMEs for which the helicity injection was monotonically increasing but followed by a sign reversal), the correlation coefficient for CME speed and $|\dot{H_{avg}}|$ was found to be -0.76 and for CME speed and $\phi(t_f)$ it was 0.77. Although the correlation coefficients are high for group B events, they are not statistically significant (as the number of events is minimal for group B). \par

\citet{2005ApJ...634L.121Q} studied $\phi_{RC}$ of 13 CME source regions of varying magnetic configurations and found a strong correlation (with a linear correlation coefficient of 0.89 at greater than 99.5\% confidence level) between CME plane-of-sky speeds and associated $\phi_{RC}$. The study also suggested that the kinematics of CMEs is probably independent of magnetic configurations of their sources. \citet{2009A&A...499..893M} combined $\phi_{RC}$ and linear speed of five CME events analyzed in their study with those from the other events, derived by \citet{2005ApJ...634L.121Q}, \citet{2007ApJ...659..758Q}, and \citet{2007SoPh..244...45L} and found a significant correlation (r= 0.76) with a confidence level greater than 99\%. Our result confirms both \citet{2005ApJ...634L.121Q} and \citet{2009A&A...499..893M} with better statistics. In our study, the linear correlation coefficient between $\phi_{RC}$ and $V_{gcs}$ is found as 0.66 (99.99\%). The accuracy of our findings is expected to be better as we consider the deprojected speed of CMEs and vector magnetograms of ARs to calculate the $\phi_{RC}$ of CME sources. The mean relative error for $\phi_{RC}$  is estimated from the average error of $\phi_{AR}$ over the pixels above the noise level. The error is inferred to be 5\% in our dataset. We also consider the uncertainty in $V_{gcs}$. \citet{2009SoPh..256..111T} found that the mean uncertainty involved in obtaining the height of CME using the GCS model is 0.48 Rs. We consider this uncertainty into the linear fitting process to estimate the error involved in $V_{gcs}$ calculation. We find a mean relative error of 12.4\% for the $V_{gcs}$ of our events. The estimated error is quite similar to what \citet{2013JGRA..118.6858S} found in measuring the deprojected propagation speed of 86 full Halo CMEs using the GCS model.

A recent study by \citet{GOPALSWAMY2017b} found a significant positive correlation ($r$= 0.6 at 99.99\%) between the speed of 48 CMEs that have signatures in interplanetary medium (in the form of magnetic clouds and non-cloud ejectas) and associated $\phi_{RC}$s. It must be noted that for the study they used the \citet{0004-637X-652-2-1740} flux rope model and deprojected speed of CMEs from the flux rope fit. They used CME observations from a single view (SOHO/LASCO) compared to the multi-view observations used in our study. In Fig. 8, we compare the reconnection flux-CME speed relation between the events of solar cycle 23 and 24. The reconnection flux and CME speed information of the events of cycle 23 are taken from \citet{GOPALSWAMY2017b}. The filled blue symbols in the figure represent the events of cycle 24. We find similar slopes for both the regression lines representing the linear least square fits CME speed-reconnection flux pairs of the events of two different solar cycles. We combine the events of both the solar cycles and find the regression equation of the linear least square fit to the scatter plot of total 81 events (the associated regression line is shown in red colour in Figure 8). The relationship established from this combined and more statistically significant database is

\begin{equation}
V_{cme}= 355 \phi_{RC}^{p\ 0.69} {km} {s^{-1}},
\end{equation}
Here $V_{cme}$ stands for the deprojected CME speed estimated from both the single view and multi-view observations and $\phi_{RC}$ is in $10^{21}$ Mx unit. The power-law relationship between $\phi_{RC}$ and $V_{cme}$ depicted in Equation 10 has an exponent $\approx 0.7$. We note that \citet{2016AN....337.1002V} found a linear relationship between peak velocity of the eruption and the added flux to the erupting flux rope by the reconnection process. 
\par
\begin{figure}
\centering
  \includegraphics[width=0.5\textwidth]{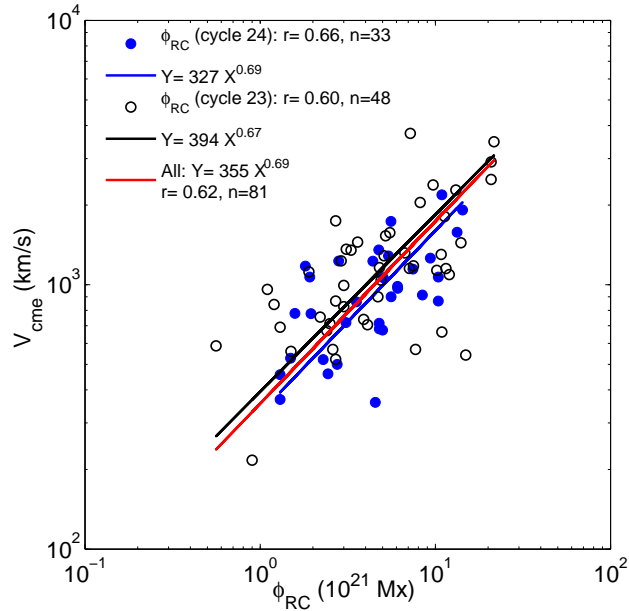}
   \caption{Scatter plot between CME speed and $\phi_{RC}$ of 33  events of solar cycle 24 and 48 events of solar cycle 23. Solar cycle 23 data is acquired from \citet{GOPALSWAMY2017b}. The filled blue symbols represent the events of cycle 23. The black, blue and continuous red lines are the regression lines derived from least-squares fits to the scatter plot of the events of cycle 23, 24 and combined cycle 23-24 data, respectively. The corresponding regression line equations are depicted in the figure.}
\end{figure}

We also find a significant positive correlation ($r\approx 0.6$ at 99\% confidence level) between CME kinematics (i.e. speed and kinetic energy) and some of the magnetic properties of CMEs (i.e., magnetic field intensity, magnetic pressure, and magnetic helicity) at 10 $R_{s}$. \citet{GOPALSWAMY2017b} studied the relationship between CME speed and its magnetic field intensity at 10 $R_s$ for 48 CMEs and found a positive correlation with $r$= 0.58 (at 99.9\% confidence level), which is similar to what we find. We study two additional magnetic parameters of CMEs (i.e., magnetic pressure and magnetic helicity) and find a good positive correlation between the parameters and the CME kinematics with a correlation coefficient of $\approx 0.64$ at 99\% confidence level.   \par

\section{Conclusion}

In this study, we obtain the deprojected physical parameters of flux rope CMEs of solar cycle 24 and calculate their magnetic (azimuthal flux, axial magnetic field intensity, and magnetic helicity) and kinetic parameters (speed and kinetic energy). Next, we measure the magnetic parameters of the associated source ARs and find the dependency of near-Sun CME kinematics on the AR magnetic parameters. We explain the basis of the relationship found between these parameters and also investigate the correspondence between the magnetic and kinetic properties of CMEs. The main conclusions of this study are:

\begin{enumerate}
\item The area and nonpotentiality of the entire source regions and the speed of associated CMEs are weakly correlated. The reason is probably the small average ratio ($\approx 0.3$) of reconnection flux during eruptions and the total flux in the source ARs. The smaller value of the flux ratio indicates that usually only a fraction of an AR involves an associated eruption.

\item The flare reconnection flux is a proxy of the reconnection energy associated with an eruptive event. In our study, we find a good correlation between CME kinematics (speed and kinetic energy) and reconnection flux with $r$= 0.66 and 0.68 in case of CME speed and kinetic energy, respectively. The slope of the regression line fitted to the reconnection flux-CME speed pairs for the events of solar cycle 24 is 0.69 which is in agreement with that derived by \citet{GOPALSWAMY2017b} for the events of solar cycle 23. The regression equation for the combined 81 events of both cycle 23 and 24 can be further used as an empirical model for predicting the near-Sun speed of CMEs.

\item The magnetic content of a CME flux rope is well correlated with its velocity and kinetic energy. We find a good correlation between the magnetic pressure of CME and its kinetic energy. This relationship is evident from the fact that the rapid expansion of CME occurs due to the higher magnetic pressure of CME flux rope relative to that of the background magnetic field.

\item We find that CME speed increases with the coronal magnetic helicity carried by the CME flux rope.
\end{enumerate}

\acknowledgments

The Center of Excellence in Space Sciences India (CESSI) is funded by the Ministry of Human Resource Development, Government of India. We acknowledge the use of data from RibbonDB catalog and \citet{GOPALSWAMY2017b}. This work benefitted from interactions mediated by an AOARD grant. The work of N. Gopalswamy was supported by the NASA's Living with a Star program. We are thankful to the U.S Naval Research Laboratory (NRL), CDAW Data Center and STEREO Science Center (SSC) for making available publicly the LASCO and STEREO databases. We thank the AIA and HMI teams for providing us with the SDO/AIA and SDO/HMI data.

\end{document}